\RequirePackage[2020/01/01]{latexrelease}

\documentclass[preprint2]{aastex631}

\usepackage{amsmath}
\usepackage{nicefrac}

\newcommand{\be}{\begin{equation}}
\newcommand{\ee}{\end{equation}}

\newcommand{\ba}{\begin{eqnarray}}
\newcommand{\ea}{\end{eqnarray}}

\newcommand{\ve}{\varepsilon}

\shorttitle{FRBs generated at the reverse shock of a powerful magnetar flare}
\shortauthors{Khangulyan et al.}
\graphicspath{{./}{figures/}}

\begin{document}

\title{Fast radio bursts by high-frequency synchrotron maser emission generated at the reverse shock of a powerful magnetar flare}

\correspondingauthor{Dmitry Khangulyan}
\email{d.khangulyan@rikkyo.ac.jp}
\author[0000-0002-7576-7869]{D. Khangulyan}
\affiliation{Graduate School of Artificial Intelligence and Science, Rikkyo University  \\ 3-34-1 Nishi-Ikebukuro, Toshima-ku, Tokyo 171-8501, Japan}

\author[0000-0002-0960-5407]{Maxim~V. Barkov}%
\affiliation{Institute of Astronomy, Russian Academy of Sciences, Moscow, 119017 Russia}

\author[0000-0002-4292-8638]{S.B. Popov}
\affiliation{Department of Physics, Lomonosov Moscow State University, Moscow, 119991 Russia}
\affiliation{Sternberg Astronomical Institute, Lomonosov Moscow State University, Moscow, 119234 Russia}



\begin{abstract}
  We consider a magnetar flare model for fast radio bursts (FRBs). We show that
  millisecond burst of sufficient power can be generated by synchrotron maser
  emission ignited at the reverse shock propagating  through the weakly magnetized 
  material that forms the magnetar flare. If the maser emission is generated in an anisotropic regime (due
  to the geometry of the production region or presence of an intense external
  source of stimulating photons) the duration of the maser flashes is similar to
  the magnetar flare duration even if the shock front radius is large. Our
  scenario allows relaxing the requirements for several key parameters: the
  magnetic field strength at the production site, luminosity of the flare, and
  the production site bulk Lorentz factor. To check the feasibility of this
  model, we study the statistical relation between powerful magnetar flares and
  the rate of FRBs. The expected ratio is derived by convoluting the
  redshift-dependent magnetar density with their flare luminosity function above
  the energy limit determined by the FRB detection threshold. We obtain that
  only a small fraction, \(\sim10^{-5}\), of powerful magnetar flares trigger
  FRBs. This ratio agrees surprisingly well with our estimates: we obtained that
  \(10\%\) of magnetars should be in the evolutionary phase suitable for the
  production of FRBs, and only \(10^{-4}\) of all flares are expected to be
  weakly magnetized, which is a necessary condition for the high-frequency maser
  emission.
\end{abstract}



\section{Introduction} \label{sec:intro}
Astrophysical relativistic shocks are a prominent site for production of broadband emission because of their apparent
ability to accelerate non-thermal particles very efficiently. Another, potentially very important, feature of
relativistic shocks is that they boost to relativistic energies the minimum energy of the particles in the
downstream \citep[see ][]{2017ApJ...835..248W}. Thus, particle distributions with inversely populated energy levels in the relativistic domain can be created. This opens the possibility for the operation of radiation mechanisms involving stimulated
emission in the radio band. If such an emission is detected, it
should allow an accurate diagnostic of the physical conditions, such as particle density and plasma magnetization, in
the environment created by relativistic shocks.

Among other situations, recently relativistic shocks were applied to explain observational properties of fast radio
bursts (FRBs). These millisecond-scale transient events were discovered by \cite{2007Sci...318..777L} (see a brief
recent review in \citealt{2020Natur.587...45Z}).  Presently, several hundred non-repeating events have been
reported\footnote{See on-line data at https://www.herta-experiment.org/frbstats/catalogue}, and several tens of
repeating sources are known, from some of which tens and even hundreds bursts were detected.\footnote{See,
  e.g. https://www.chime-frb.ca/repeaters for the CHIME telescope data on repeating sources.} At the moment, bursts
themselves are detected only in radio at frequences from $\sim100$~MHz up to $\sim10$~GHz (see
\citealt{2021Univ....7...76N} about multiwavelength observations of bursts and their sources).

Several observational features of FRBs suggest that a coherent radiation mechanism is responsible for their generation \citep{2014MNRAS.442L...9L,2014PhRvD..89j3009K}.
In particular, these coherent emission ``smoking guns'' include high luminosity and nearly \(100\%\) linear polarization
detected for some FRBs \citep[see, e.g.,][]{2018ApJ...863....2G,2018Natur.553..182M,2019MNRAS.488..868O}.  This favors scenarios 
involving synchrotron maser emission, however the specific realization of the process sill remains debated. Synchrotron maser 
emission can be produced at relativistic
gyro frequency \citep[see, e.g., in][note that in the case of magnetized plasma, relativistic gyro frequency and plasma
frequency have similar values]{1991PhFlB...3..818H,1992ApJ...391...73G,2019MNRAS.485.3816P}. However, for weakly
magnetized shocks synchrotron maser emission can be generated also at significantly higher frequencies \citep[see, e.g.,][]{1970SvA....13..797S}. We apply this
scenario to estimate the frequency of synchrotron maser emission behind pulsar wind termination shock (TS) and at
relativistic shocks caused by magnetar flares.

Magnetar bursts were proposed as possible sources of FRBs already in 2007 (see \citealt{2010vaoa.conf..129P}). Now,
leading scenarios of FRB activity are related to this type of neutron stars (see a review in
\citealt{2021Univ....7...56L}). Magnetars are neutron stars with strong magnetic fields (see a review e.g. in
\citealt{2015RPPh...78k6901T}). In the first place, they are known as sources of powerful bursts with total luminosity
covering a wide range up to $\sim10^{47}$~erg~s$^{-1}$. Strong bursts are rare following a power-law distribution
$dN/dE_\mathrm{fl}\sim E_\mathrm{fl}^{-\gamma}$ with $\gamma\approx(1.4-2)$, here $E_\mathrm{fl}$ is the total energy of
the flare. Three most energetic bursts --- so-called \emph{giant flares} and/or \emph{hyper flares}, --- were detected from Galactic
sources. However, several well-established candidates for extragalactic flares are also known (see
e.g. \citealt{2021ApJ...907L..28B} and references therein). In 2020 simultaneous bursts in radio
\citep{2020Natur.587...54C, 2020Natur.587...59B} and X/$\gamma$-rays \citep{2020ApJ...898L..29M, 2021NatAs...5..372R, 2021NatAs...5..401T, 2021NatAs.tmp...54L} were detected from a Galactic magnetar SGR 1935 (Soft Gamma Repeater). This made links between FRB sources and magnetars even stronger.

In contrast to other studies \citep[see, e.g.,][]{2017ApJ...842...34W} we do not adopt assumptions regarding the isotropy  of the stimulated emission in the plasma co-moving frame and consider a possibility that the maser emission features a significant anisotropy (detailed analysis will be presented in Khangulyan et al. 2022).  We show that anisotropy of the maser emission would imply a short duration of the maser flashes triggered by magnetar flares.  Furthermore, our estimates show that the conditions, which can be naturally achieved at powerful
magnetar flares, are sufficient for generation of synchrotron maser emission in the GHz band. This supports the
scenarios that suggest magnetars as sources for FRBs \citep[see, e.g.,][]{2020ApJ...896..142B, 2017MNRAS.468.2726K,
  2014MNRAS.442L...9L, 2020ApJ...897....1L, 2019MNRAS.485.4091M, 2016ApJ...824L..18L, 2017ApJ...838L..13L, 2021arXiv210207010L} and alleviate the extreme assumptions required for
their realizations.  Moreover, if a magnetar is located in a binary system or moves with high proper speed through the
interstellar medium, this can even further increase the frequency at which synchrotron maser operates.
\section{Maser emission at pulsar wind termination shock}
\label{sec:magnetars}

\citet{1991PhFlB...3..818H} have shown that if electrons have a ``ring'' momentum distribution, i.e., follow
gyrorotation for several revolutions then maser synchrotron can be generated at relativistic gyrofrequency,
\be
\Omega_{L,e}=\frac{ceB}{E}\,.
\ee
Here, \(E\) and \(B\) are particle energy and magnetic field, respectively (also note that \(m_e\), \(e\), and \(c\) are the conventional constants: electron mass, elementary change, and
light speed). Particle-in-cell (PIC) simulations and theoretical considerations indicate a possible existence of several
coherent gyration cycles in the downstream of a relativistic shock
\citep{1988PhRvL..61..779L,1992ApJ...391...73G,2019MNRAS.485.3816P}. Such a particle distribution has population
inversion thus in the region with thickness $\sim r_g$ maser-synchrotron emission can be formed.  However, if the
electron distribution is less regular, then it is still unclear if maser emission can be generated in the range of
frequencies where gyrorotation is important. Moreover, the relativistic gyro frequency is typically quite low, and
extreme assumptions are required to match it to the frequencies at which FRBs are observed. For example,
\citet{2014MNRAS.442L...9L} adopted a magnetar flare magnetic field of \(B\sim10^5\rm\,G\) at the distance of
\(10^{15}\rm\,cm\), this corresponds to isotrpotic luminosity exceeding \(\gg10^{50}\rm \,erg\,s^{-1}\), which
significantly larger than the values typically assumed. Most likely, sufficiently strong magnetic field can be realized
only in the pulsar / magnetar magnetosphere \citep[see, e.g.,][]{2020ApJ...897....1L}.  Alternatively, one can assume
that the maser emission is produced at a shock that moves with large bulk Lorentz factor, \(\gg100\), in the laboratory
frame \citep{2019MNRAS.485.4091M,2019arXiv190807743B}.

Significantly above the cyclotron frequency, \(\omega\gg\Omega_{L,e}\), the dielectric permittivity is simply \(
\varepsilon = 1 - \left(\frac{\Omega_{p,e}}{\omega}\right)^2\) (for a more detailed discussion see Appendix~\ref{sec:maser}). For this regime, 
\citet{1967JETP...24..381Z} obtained that synchrotron emission is amplified if the main contribution to the absorption is provided by particles with sufficiently large energy,
\be\label{eq:maser_energy}
E >E_{\rm min}= m_ec^2  \frac{2\omega_{L,e} \omega^2}{\Omega_{p,e}^3}\,, 
\ee
where \(\omega_{L,e}={eB}/({m_e c})\) is non-relativistic cyclotron frequency.
Using this equation we can estimate the frequency below which the maser emission can be formed:
\be\label{eq:maser_energy2}
\omega<\omega_{\rm max}= \sqrt{\frac{E}{2m_ec^2}\frac{ \Omega_{p,e}^3}{\omega_{L,e}}}\,, 
\ee
which is almost identical to the expression, \(\Omega_{p,e}\sqrt{\Omega_{p,e}/\Omega_{L,e}}\), obtained by \citet{2019ApJ...875..126G}. 
At crossing of a relativistic shock, the particles get their energy boosted by a factor, \(\Gamma_{\rm sh}\), which is approximately equal to the upstream bulk Lorentz factor measured  in the frame of the shock wave. Thus, if the upstream is cold, then we can simply adopt \(E_{\rm min}\approx\Gamma_{\rm sh}m_ec^2\).

Considering the theoretical and numerical results outlined above, we may expect formation of maser emission at frequency (we refer this as ``low frequency maser window'')
\be\label{eq:maser_standard}
\omega_{\rm m,1}\sim \frac{eB}{m_ec\Gamma_{\rm sh}}\,,
\ee
from particles having a ``ring'' momentum distribution (formation of such a distribution requires a cold upstream, also see analysis with PIC simulations in \citealt{2019MNRAS.485.3816P,2020MNRAS.499.2884B}); and at (we referee this as ``high frequency maser window'')
\be\label{eq:condition}
\Omega_{L,e}\ll\omega_{\rm m,2}< \sqrt{\frac{\Omega_{\rm p,e}^3}{2\Omega_{\rm L,e}}}\,.
\ee
The latter regime can be realized only if
\be
\frac{\Omega_{L,e}}{\Omega_{\rm p,e}}\ll 1\,.
\ee
This condition can be rewritten as
\be
\frac{B^2/(4\pi)}{n_e \Gamma_{\rm sh} m_ec^2}\ll1\,,
\ee
which implies a condition on the plasma magnetization (i.e., the ratio of the Poynting flux to the plasma kinetic energy flux) in the upstream of the shock:
\(\sigma\ll1\)
(where we ignore a factor of \(\sim3\) for simplicity). This is consistent with previous analysis of this process: high-frequency maser emission can be generated in weakly magnetized plasma \citep{1970SvA....13..797S,2002ApJ...574..861S,2019ApJ...875..126G}.

The
shock magnetization of pulsar winds  might be quite small, \(10^{-3\dots-1}\). Thus, a priory, we cannot exclude that the conditions behind astrophysical shocks, in particular, 
pulsar wind TSs, are suitable for production of synchrotron maser emission in the range of frequencies
\(\omega_{\rm m,2}\). In what follows we estimate the frequency that corresponds to the high-end of the range
\be
\omega_{\rm max} =\sqrt{\frac{E_{\rm min}\Omega_{\rm p,e}^3}{2m_ec^2\omega_{\rm L,e}}}\,.
\ee
According to Eq.~\eqref{eq:condition}, maser emission can be generated in the range  \( \sigma^{\nicefrac{3}{4}}\omega_{\rm max} \ll\omega<\omega_{\rm max}\). To compute the actual absorption coefficient, one needs to know the electron distribution and then to take the integration in Eq.~\eqref{eq:absorption}. Figure~1 in \citet{2019ApJ...875..126G} shows that the frequency range with negative absorption coefficient is quite narrow, between \(\omega_{\rm max}/3\) and \(\omega_{\rm max}\).

Generation of maser emission at relativistic gyrofrequency, Eq.\eqref{eq:maser_standard}, is discussed in a number of
papers including its implication for FRBs \citep{2014MNRAS.442L...9L,2020ApJ...897....1L,2019MNRAS.485.4091M, 2020ApJ...896..142B}. The possibility of production
of FRBs by synchrotron maser emission in the range given by Eq.~\eqref{eq:condition} got much less attention. However,
this range has an obvious advantage --- this mechanism allows producing coherent emission at significantly higher
frequencies, thus it can alleviate the need for extreme assumption adopted, e.g., in \citet{2014MNRAS.442L...9L}.  Below we discuss the conditions required for its ignition at a pulsar (or a magnetar) wind
TSs. We consider two cases: the TS formed by a steady pulsar wind (``steady case'') and
the interaction of an intense flare with pulsar wind nebula (``non-steady case'').

\subsection{ Steady case}
Conditions behind a steady reverse shock in a pulsar wind are determined by a few parameters: the pulsar spin-down
luminosity, \(L_{\rm sd}\), the pulsar wind magnetization, \(\sigma\), its bulk Lorentz factor, \(\Gamma_{\rm wind}\), and
the radius of the TS, \(R_{\rm ts}\). As we are interested in the case with \(\sigma\ll1\) and the upstream
bulk Lorentz factor is large, the downstream speed is simply \(\nicefrac{c\,}{3}\) (the bulk Lorentz factor is \(\nicefrac{3}{\sqrt{8}}\)). This allows obtaining all other parameters of the downstream. Namely, the magnetic field (in the plasma co-moving frame) is
\be\label{eq:B_steady}
B\approx \sqrt{\frac{8\sigma L_{\rm sd}}{R_{\rm ts}^2 c}}\,;
\ee
electron number density (in the plasma co-moving frame)
\be
n_e\approx \frac{(1-\sigma)L_{\rm sd}}{\sqrt{2}\pi R_{\rm ts}^2 m_ec^3\Gamma_{\rm wind}}\,;
\ee
plasma internal energy
\be
\ve_{\rm pwn}\approx \frac{(1-\sigma)L_{\rm sd}}{\sqrt{2}\pi R_{\rm ts}^2 c}\,.
\ee

Thus, we obtain that the plasma and cyclotron frequencies are
\be
\Omega_{\rm p,e} = 2^{\nicefrac{3}{4}}\frac{e}{m_ec}\sqrt{\frac{(1-\sigma)L_{\rm sd}}{ R_{\rm ts}^2 c\Gamma_{\rm wind}^2}}\,
\ee
and 
\be
\omega_{\rm L,e} = \frac{e}{m_ec}\sqrt{\frac{8\sigma L_{\rm sd}}{R_{\rm ts}^2 c}}\,.
\ee
Thus, the maximum frequency for the synchrotron maser radiation is 
\be
\begin{split}
  \omega_{\rm max} &\approx 2^{\nicefrac{-1}{8}}\frac{e}{m_ec} \sqrt{\frac{L_{\rm sd}}{R_{\rm ts}^2 c}} \frac{(1-\sigma)^{\nicefrac{3}{4}}}{ \Gamma_{\rm wind}\sigma^{\nicefrac{1}{4}}}\,\\
  &\approx 3\times10^3L_{\rm sd,38}^{\nicefrac{1}{2}}R_{\rm ts,15}^{-1}{ \Gamma_{\rm wind,3}^{-1}\sigma_{-2}^{\nicefrac{-1}{4}}}\quad[{\rm rad \; s^{-1}}]\,,
\end{split}
\label{eq:omax}
\ee
where \(L_{\rm sd} =  10^{38}L_{\rm sd, 38} \rm\,erg\,s^{-1}\), and we adopted \(E_{\rm min} = m_ec^2\Gamma_{\rm wind}\).

As we can see, even for the ``generous'' values used for the normalization in Eq.~\eqref{eq:omax}, at a relativistic shock formed by a steady pulsar wind, the  maser emission can be generated at very low frequencies only. 

\subsection{Non-stationary case}

If a powerful flare hits a standing shock (which is assumed to be the TS of the wind), then a system of two relativistic shocks is to be formed. The forward shock (FS) 
propagates through the matter in the nebula, and the reverse shock (RS) through the material that forms the flare.  In the
laboratory frame both shocks (and also the contact discontinuity --- CD hereafter) can move with relativistic speed. To estimate
these speeds one needs to consider the jump condition at each shock and pressure balance at the CD.

Dynamics of the FS and RS is discussed in the Appendix~\ref{sec:MHD}, we just adopt two key results from there \citep[for a discussion in detail, see][]{1976PhFl...19.1130B}. The bulk Lorentz factor of the shocks
\be
\Gamma_{\rm fs}\approx\Gamma_{\rm rs}\approx \Gamma \approx \frac12\sqrt[4]{\frac{L_{\rm fl}}{L_{\rm sd}}}\,;
\label{eq:GGG}
\ee
and flare penetration distance to the PWN:
\be
\Delta R \approx \Delta t_{\rm fl}c\sqrt{\frac{L_{\rm fl} }{L_{\rm sd}}}\,.
\ee
The typical energy associated with FRBs is \(\sim10^{40}\rm\,erg\), since maser mechanism can radiate away a per-cent fraction of energy \citep[see][and reference therein]{Koryagin2000}. Thus, it is feasible that FRBs require magnetar flare of energy \(\sim10^{42}\rm\,erg\) and luminosity \(L_{\rm fl}\sim10^{45}\rm\,erg\,s^{-1}\) (given the ms duration). This value is significantly smaller than the maximum recorded flare luminosity (see in Sec.~\ref{sec:intro}). 

If the wind magnetization is small, the magnetic field close to the TS is small, given by Eq.~\eqref{eq:B_steady}. In the frame of the FS, the strength of the magnetic field is amplified by a factor of \(\Gamma_{\rm fs}\), but the flow magnetization remains small, since the plasma internal energy is also amplified by the same factor. Thus, the conditions at the FS of flare should remain suitable for production of the maser emission independently on the flare magnetization. Relativistic cyclotron, \(\Omega_{L,e}\), and plasma, \(\Omega_{p,e}\), frequencies (notice the capital letter notation in contrast to small letters for the non-relativistic case) do not change by the compression by the forward shock, thus the maser frequency in the FS downstream is given by Eq.~\eqref{eq:omax}.  We therefore need to account only for the Doppler boosting:
\be
  \omega_{\rm max, fs}=  2\omega_{\rm max}\Gamma_{\rm fs}\,.
\ee
Substituting Eq.~\eqref{eq:GGG} to Eq.~\eqref{eq:omax} we obtain 
\be
\label{eq:omaxfs}
\begin{split}
  \omega_{\rm max, fs}\approx & 2^{\nicefrac{-1}{8}}\frac{e}{m_ec} \sqrt{\frac{L_{\rm sd}^{\nicefrac{1}{2}}L_{\rm fl}^{\nicefrac{1}{2}}}{R_{\rm ts}^2 c}} \frac{(1-\sigma)^{\nicefrac{3}{4}}}{ \Gamma_{\rm wind}\sigma^{\nicefrac{1}{4}}}\,\\
  \approx& 3\times10^4\,[{\rm rad \; s^{-1}}]\quad\times\\
  &L_{\rm sd,35}^{\nicefrac{1}{4}}L_{\rm fl,45}^{\nicefrac{1}{4}}R_{\rm ts,15}^{-1}{ \Gamma_{\rm wind,3}^{-1}\sigma_{-2}^{\nicefrac{-1}{4}}}\,,
\end{split}
\ee

If we consider in the RS frame the formation of the maser emission at the RS, is identical to the emission at the pulsar wind TS. Thus, we should replace \(L_{\rm sd}\) with \(L_{\rm fl}/ {4}\Gamma_{\rm rs}^2\) and \(\Gamma_{\rm wind}\) with \(\Gamma_{\rm fl}/{2}\Gamma_{\rm rs}\) in Eq.~\eqref{eq:omax} and account for the Doppler boosting.  We therefore obtain 
\be\label{eq:omaxrs}
\begin{split}
  \omega_{\rm max,rs}\approx & 2^{\nicefrac{-1}{8}}\frac{e}{m_ec^{\nicefrac{3}{2}}} \frac{L_{\rm fl}^{\nicefrac{3}{4}}}{L_{\rm sd}^{\nicefrac{1}{4}} R_{\rm ts}} \frac{(1-\sigma_{\rm fl})^{\nicefrac{3}{4}}}{ \Gamma_{\rm fl}\sigma_{\rm fl}^{\nicefrac{1}{4}}}\,\\
  \approx & 3\times10^{9}\,[{\rm rad \; s^{-1}}]\quad\times\\
  &L_{\rm sd,35}^{\nicefrac{-1}{4}} L_{\rm fl,45}^{\nicefrac{3}{4}} R_{\rm ts,15}^{-1}{ \Gamma_{\rm fl,3}^{-1}\sigma_{\rm fl,-2}^{\nicefrac{-1}{4}}}\,,
\end{split}
\ee
here \(\sigma_{\rm fl}\) is magnetization of the flare.

Our estimates, Eqs.~(\ref{eq:omax},\ref{eq:omaxfs}) and \eqref{eq:omaxrs}, for the frequency at which synchrotron maser emission can be generated,  show that in the case of the TS formed by a steady pulsar wind, given by Eq.~\eqref{eq:omax}, the
maser emission appears in the kHz. Thus, it remains undetectable even if we adopt extreme assumptions regarding the
pulsar wind luminosity and the shock formation distance. In contrast, in the case of the shocks created by intense magnetar flares at RS, given by Eq.~\eqref{eq:omaxrs}, the maser frequency can reach the GHz band without invoking any extreme assumptions. In what follows we mostly focus at the maser emission generated at the RS of the magnetar flare. 

\section{Application to the magnetar scenario for  FRBs}\label{sec:application}

The frequency of maser radiation produced at the RS of a magnetar flare, Eq.~\eqref{eq:omaxrs}, is determined mostly by
the radius of the standing shock and luminosity of the flare,
$\omega_{\rm max} \propto R_{\rm ts}^{-1}L_{\rm fl}^{\nicefrac{3}{4}}$.  If this mechanism is responsible for FRBs, which are
detected in the GHz band, it requires either $R_{\rm ts}< 10^{15}$~cm or \(L_{\rm fl}\gtrsim 10^{45}\rm\, erg\,s^{-1}\). The isotropic luminosity of magnetar flares achieves, in some cases, \(\sim10^{47}\rm\, erg\,s^{-1}\) (please, see the Sec.~\ref{sec:intro}), and the required luminosity of \(10^{45}\rm\,erg\,s^{-1}\) seems to be reasonable. However, given the stronger dependence on the TS radius, below we check if the used normalization of \(10^{15}\rm\,cm\) is reasonable.

For isolated magnetars, in a very rough way, the TS radius, \(R_{\rm ts}\), is determined by the external pressure, \(p_{\rm ext}\): 
\be 
R_{\rm ts} = \sqrt{\frac{L_{\rm sd}}{4\pi c p_{\rm ext}}}\sim10^{15} L_{\rm sd,34}^{1/2} p_{\rm ext,-8}^{-1/2}\rm\,cm\,. 
\label{eq:rts}
\ee
Here we accounted that given the magentar typical rotation period of a few seconds, the spin-down power is very modest, 
\(L_{\rm sd}\sim 10^{34}\rm\,erg\,s^{-1}\), and the external pressure can be normalized to \(10^{-8}\rm\,dyn\,cm^{-2}\)
\citep[see, e.g.,][for a discussion]{2014MNRAS.442L...9L}. Thus, it seems quite feasible that the TS in
nebula formed by magnetar wind is \(R_{\rm ts}\sim10^{15}\rm \,cm\). There are, however, two processes that can increase the
magnetar wind TS radius: (i) the energy injection by the flares and (ii) and pressure drop inside
supernovae (SN) during the phase of the adiabatic expansion or late Sedov phase. Below we briefly check if these effects impose any
significant constraints.

As we can see from Eq.~\eqref{eq:dr}, when the TS is hit by a flare, its position is displaced by \(\Delta R\). If the shock recovery time $t_{\rm rec} \sim 3 \Delta R/ c \sim 3 \Delta t_{\rm fl}\sqrt{L_{\rm fl}/L_{\rm sd, eff}}$ is long compared to the delay between flares, $T_{\rm fl}$, then  the shock position is determined by the effective magnetar ``spindown'' luminosity, which accounts for the energy injection by the flares:
\be
\begin{split}\label{eq:lsdef}
  L_{\rm sd, eff} &= L_{\rm sd} + E_{\rm fl}/T_{\rm fl} \\
  &\approx 10^{35}(E_{\rm fl,42} T_{\rm fl, 7}^{-1}+0.1L_{\rm sd,34})\, \rm erg\,s^{-1}\,.
\end{split}
\ee
To account for this effect, below we use $L_{\rm sd, eff} $ instead of $ L_{\rm sd}$.

If a magnetar is located inside a SN remnant (SNR), then the radius of the nebula (and of the TS) is
determined by the pressure dynamics in the center of the SNR. During the first several hundred years, the SN shell rapidly 
expands during the ejecta dominated phase and the RS does not reach the center of the SNR. Low pressure there
should allow almost a free nebula expansion, and the radius of the magnetar wind TS can be very large,
\(R_{\rm ts}\sim10^{17}\rm\,cm\), even if the spindown losses are small. After approximately \(10^3\rm\,yr\), the
explosion enters the Sedov phase, the expansion slows down and the  RS reaches the SNR center. This
compresses the nebula and establishes the magnetar wind TS at 
\be
R_{\rm ts} \approx  10^{15} L_{\rm sd,35}^{1/2} t_{10.5}^{3/5}\quad {\rm cm}\,,
\label{eq:tsrss}
\ee
where \(t\) is the time elapsed since SN explosion \citep[see, e.g.,][]{2014ApJ...785..130Z}. Here we ignore the
magnetar braking, which significantly decrease the radius of the TS after \(10\rm\,kyr\) \citep[see,
e.g.,][]{2018ApJ...860...59K}, as it is uncertain if magnetars are capable to produce frequent powerful flares at their late
evolution phase. Thus, being conservative we adopt that for magnetars reside insider a SNR, the time span during
which the radius the wind TS is limited to \(10^{15}\rm \,cm\) is about \(3\)~kyr. {By studying the properties of the persistent radio source associated with FRB 121102 \citep{2017Natur.541...58C,2017ApJ...834L...8M} \citet{2017ApJ...842...34W} derived an upper limit of \(10^{2.5}\rm\,yr\) on the age of the source. This estimate seems to be consistent with the source age allowed in the framework of our model. We note, however, that the estimate by \citet{2017ApJ...842...34W} is obtained under the assumption of smoothly changing conditions in the radio source, thus in the context of our scenario this age limit should be considered as an upper limit on the time elapsed since the nebula got compressed by the reverse shock.}

To escape from SNR, a large magnetar proper speed, \(v\), is required. In this case, the interaction with the interstellar medium of density \(\rho_{\rm ism}\)
creates a bow shock at
\be\label{eq:rbs}
\begin{split}
  R_{\rm ts, bow} &= A\sqrt{\frac{L_{\rm sd} }{ 4 \pi c  \rho_{\rm ism} v^2}}\,,\\
  &\approx 4 \times 10^{14} A_{-0.5} L_{\rm sd, 35}^{1/2} \, n_{\rm ism, 1}^{-1/2} v_{8}^{-1} \, {\rm cm}\,.
\end{split}
\ee
The factor
$A = R_{\rm rs}/ R_{\rm fs}\sim 1/3$ accounts for the ratio of the RS to FS distances in bow shock nebulae
\citep{2019MNRAS.484.4760B}. Finally, we note that if a magnetar is located in a binary system \citep[see][for
observational hints for binary systems harboring magnetars]{2020PhRvL.125k1103Y}, then the shock locates at distances
comparable to the orbital separation and it can be very small, \(10^{12}\rm\,cm\). However, because of severe free-free
absorption in the circumbinary environment, FRBs can be generated in binary systems with intermediate star separations, \(\sim10^{13}\rm cm\) or more \citep{2020ApJ...893L..39L}. 

We therefore conclude that magnetar flares of the intermediate  luminosity of \(\sim10^{45}\rm \,erg\,s^{-1}\) can generate
synchrotron maser bursts in the GHz energy band in the nebulae around (i) isolated magnetars during several kyr
of their evolution; (ii) in run-away magnetars moving with high proper speed; and (iii) in magnetars in binary systems with
orbital separation of \(< 10^{14}\rm\,cm\). This implies that in \(\sim10\%\) nebulae around active magnetars, the
standing shock is at the distance suitable for production of FRBs.
\section{Time profile of the signal emitted by a maser in the relativistic blast wave}\label{ap:boosting}

If in the plasma co-moving frame the produced emission is isotropic, then in the observer frame because of photon
aberration it is focused into a beaming cone with opening angle of \(\Gamma^{-1}\). Then a characteristic time-scale of
\(R / (\Gamma^2 c)\) determines the shortest duration of a pulse produced by a spherical blast wave. Here \(R\) and
\(\Gamma\) are the blast wave radius and bulk Lorentz factor. Thus, for the typical ms duration of FRBs implies the
shock radius of \(R\ll 10^{8}\Gamma^2 (\Delta t/{\rm ms}) \rm \,cm\).

However, there are several physical mechanisms that could lead to an anisotropic maser emission. First of all, if an
external source provides sufficiently intense field of stimulating photons then the produced maser emission should be
predominately directed away from this source. It is natural to expect that such a dominant source may determine the
preferred direction for the maser emission if the maser emission production site has a quasi spherical shape. In the
opposite case, when the production site is significantly smaller in one of the directions, then the intensity of locally
generated emission might be highest along the source largest extension direction. Thus, 
again this emerges into a strongly anisotropic direction diagram of the maser emission. For example, one may expect realization of such
a scenario if the maser emission is generated in a thin shell (detailed analysis will be presented
in Khangulyan et al. 2022).

\begin{figure}
  \plotone{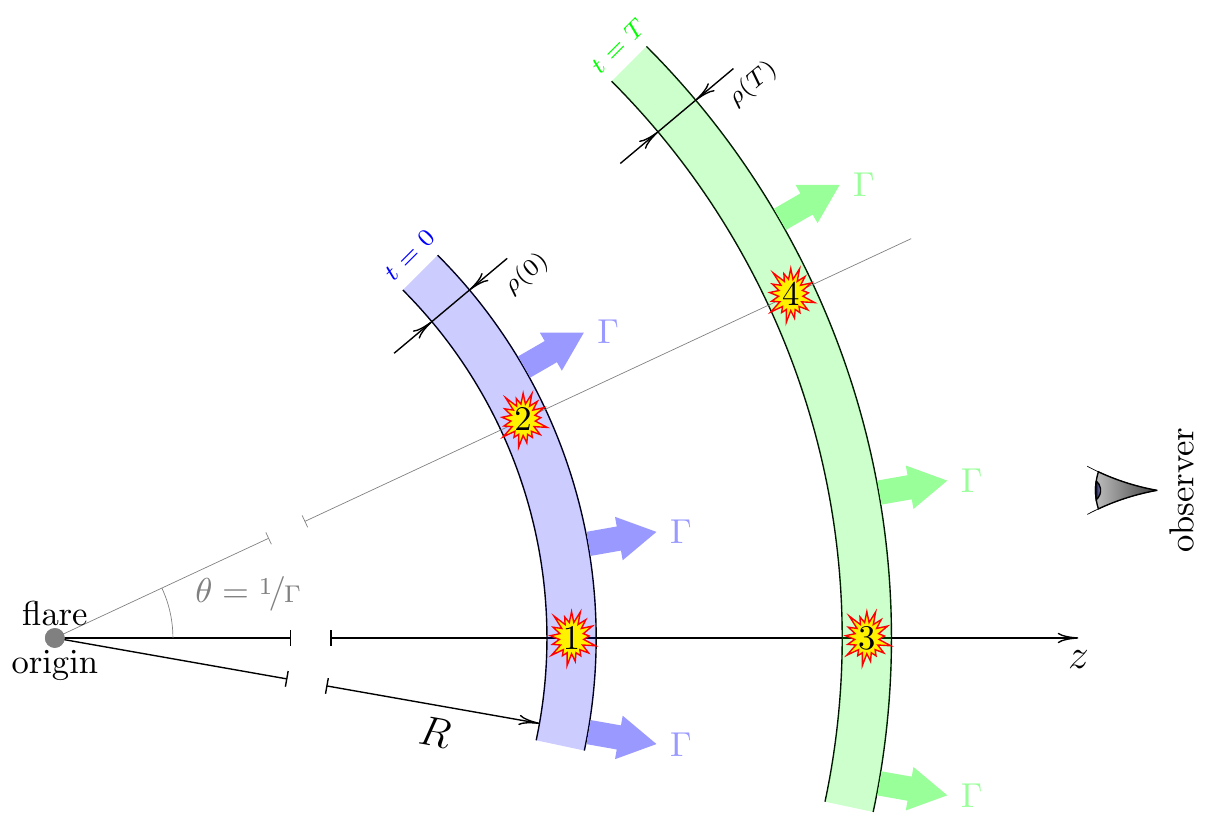}
  \caption{Depending on the anisotropy of the emission, the observer registers emission components produced in different parts of the shell. This determines the apparent signal duration. \label{fig:variability}}
\end{figure}

Figure~\ref{fig:variability} presents a sketch which illustrate how the signal duration depends on the anistorpy of the emission. It is assumed that an emitting shell has thickness, \(\rho(t)\), and radius, \(R(t)\),  and expands with speed, \(v\). The shell is assumed to be thin,   \(\rho(t)\ll R(t)\), and its expansion speed to be relativistic, \(\Gamma=1/\sqrt{1-(v/c)^2}\gg1\). The emission process starts at a time instant \(t=0\) and terminates at \(t=T\) (both are measured in the laboratory coordinate system). If the emission is isotropic in the plasma co-moving frame, then the observer mainly sees the emission originated in the shell patch with a typical size of \(2R/\Gamma\). The signal duration is determined by the delay between arrival times of the emission generated at the point label \(1\) and \(4\) in Fig.~\ref{fig:variability}. If the emission is strongly anisotropic in the plasma co-moving frame, then the signal duration is the ``delay'' between points labeled \(1\) and \(3\). Here we assume that the emission is generated radially away from the flare origin. If the emission is produced perpendicularly to that direction in the plasma co-moving frame, then the ``delay'' is determined by points \(2\) and \(4\). 
Simple calculations give the corresponding delays:
\be\label{eq:delays}
    \begin{matrix}
      \tau_{2}-\tau_{1}&=&\frac{R}{c}(1-\cos\theta)&\approx&\frac{R}{2c\Gamma^2}\,\\[5pt]
      \tau_{3}-\tau_{1}&=&T(1-\beta)&\approx&\frac{T}{2\Gamma^2}\,\\[5pt]
      \tau_{4}-\tau_{2}&=&T(1-\beta\cos\theta)&\approx&\frac{T}{\Gamma^2}\,.
    \end{matrix}
\ee


As it can be seen from Eq.~\eqref{eq:delays}, the signal duration is determined by the shell radius, only if the emission is isotropic in the plasma co-moving frame. In the case of anisotropic emission (somehow almost independently on the preferred angle), the signal duration depends only on the shell bulk Lorentz factor, \(\Gamma\), and its lifetime, \(T\).


For the scenario discussed here, one should use  \(T\approx\Delta R/c \approx\Delta t_{\rm fl} \sqrt{\frac{L_{\rm fl}}{L_{\rm sd}}}\), 
where \(\Delta R\) is the flare penetration distance  (see Appendix~\ref{sec:MHD}). The shock Lorentz factor 
is \(\Gamma\approx \frac12\sqrt[4]{\frac{L_{\rm fl}}{L_{\rm sd}}}\).  It implies that for the anisotropic emission case the synchrotron  
maser burst triggered by a magentar flare of duration \(\Delta t_{\rm fl}\), is seen by the observer as a flare of a similar duration 
\be
\Delta \tau \sim \Delta t_{\rm fl}\,.
\ee

Thus, we obtain that if the emission is highly anisotropic in the plasma co-moving frame, then the maser emission from the shell is to be registered during a very short time interval, comparable to the duration of the flare. {We emphasize that because of the strong anisotropy of the maser emission, the shell radius and its bulk Lorentz factor have a minor influence on the duration of the radio burst. Thus, the constraints on the shell radius and bulk Lorentz factor, which are obtained under the assumption (typically hidden) of isotropic emission in the co-moving frame, seem to be irrelevant.  }

\section{How many of Magnetars and FRBs in the Universe?}

Each magnetar undergoes many flares of different energy during its lifetime. The luminosity function for the flare energy is quite well constrained with the observations. Below we compare the expected number of the magnetar flares, which are powerful enough to produce detectable FRBs, with observational statistics of FRBs.

It is assumed that the number of magnetars is proportional to the star formation rate (SFR).
We use expression for SFR at different redshifts $z$ from \cite{2014ARA&A..52..415M}:
\begin{equation}
    \psi(z) = 0.015 \frac{(1+z)^{2.7}}{1+[(1+z)/2.9]^{5.6}} \, M_\odot\, {\rm yr}^{-1} {\rm Mpc}^{-3}.
\end{equation}

For basic cosmological equations we follow \cite{1999astro.ph..5116H}. 
For a given $z$ the comoving volume is: 
\begin{equation}
   \frac{ dV_{\rm c}}{dz}=\frac{c}{H_0}\frac{4\pi D_{\rm L}^2}{(1+z)^2\sqrt{(1+z)^3\Omega_{\rm m} + \Omega_{\Lambda}}}.
\end{equation}
Here $D_{\rm L}$ is the luminosity distance, $H_0$ --- present day Hubble constant,
$\Omega_{\rm m}$ and $\Omega_{\Lambda}$ are present day normalized matter and dark energy density (for numerical estimates we apply fiducial values 0.3 and 0.7, correspondingly). 

We assume that the Galactic SFR is 3 solar mass per year, and that there are 100 magnetars in the Milky way (i.e., about 10\% of all neutron stars younger than a few tens thousand years). 

The energy distribution of flares obeys a power-law dependence \citep{2015RPPh...78k6901T}:
\begin{equation}
    dN=A E_{\rm fl}^{-\gamma} dE_{\rm fl}.
\end{equation}
Below we use $\gamma=5/3$, and coefficient $A$ is obtained from the normalization condition: $\int_{E_{\rm min}}^{E_{\rm max}} A E_{\rm fl}^{1-\gamma}dE_{\rm fl} = 10^{48}$~erg. 
For $E_{\rm max}=10^{48}$~erg we obtain $A\approx 3 \times 10^{31}$~erg$^{2/3}$ (slightly smaller values of $E_{\rm max}$ do not change our conclusions significantly).

Total number of flares detectable from Earth from a given magnetar is limited by the energy:
\begin{equation}
   A \int_{E_{\rm lim}}^{E_{\rm max}}  E_{\rm fl}^{-5/3}dE_{\rm fl}=\frac{3}{2}\frac{A}{E_{\rm lim}^{2/3}} \approx  10^{5} E_{\rm lim,40}^{-2/3},
\end{equation}
here
\begin{equation}
E_{\rm lim}\approx 10^{40} \left(\frac{S_{\rm lim}}{0.1 \, {\rm Jy}}\right) \left(\frac{D_{\rm L}}{1\, {\rm Gpc}}\right)^2 \left(\frac{\Omega }{ 4\pi \, {\rm sr}}\right) \, {\rm erg},
\end{equation}
where $S_{\rm lim}$ is the minimum observed flux and $\Omega$ --- solid angle (we assume isotropic emission).
Here we assume that the energy emitted in radio is about 1\% of the total energy of the flare and duration of the flare is about 1 msec. Also we use expression from  \cite{2015ApJ...814L..20M} for the limiting radio luminosity.

Thus, the daily rate of flares detectable by an observer on Earth is
\begin{equation}
    \frac{10^{-7}}{1+z} \int_{E_{\rm lim}}^{E_{\rm max}} \frac{A}{ E_{\rm fl}^{\gamma}} dE_{\rm fl} {\rm \; days}^{-1} \sim \frac{10^{-2}}{(1+z) E_{\rm lim,40}^{2/3}} {\rm \; days}^{-1} .
\end{equation}
We note that the factor $1/(1+z)$ appears due to the cosmological time dilation. 
Here we assumed that magnetars are active for 30 kyr, i.e. approximately for $10^7$ day. 

So, finally the rate per day from all magnetars in the comoving volume $dV_\mathrm{c}(z)$ is:
$$
0.015 \frac{(1+z)^{2.7}}{1+[(1+z)/2.9]^{5.6}} \, M_\odot\, {\rm yr}^{-1} \, {\rm Mpc}^{-3} \times
$$
$$
\times \frac{c}{H_0} \frac{4 \pi D_{\rm L}^2}{(1+z)^2\sqrt{(1+z)^3\Omega_{\rm m}+\Omega_{\Lambda}}}  \times 
$$
$$
\times \frac{100}{3 \, M_\odot \, {\rm yr}^{-1}} \times 10^{-7} {\rm days}^{-1} \times
$$
\begin{equation}
\times \frac{1}{1+z}\int_{E_{\rm lim}}^{E_{\rm max}} A E_{\rm fl}^{-\gamma}dE_{\rm fl}.
\end{equation}

\begin{figure}
\plotone{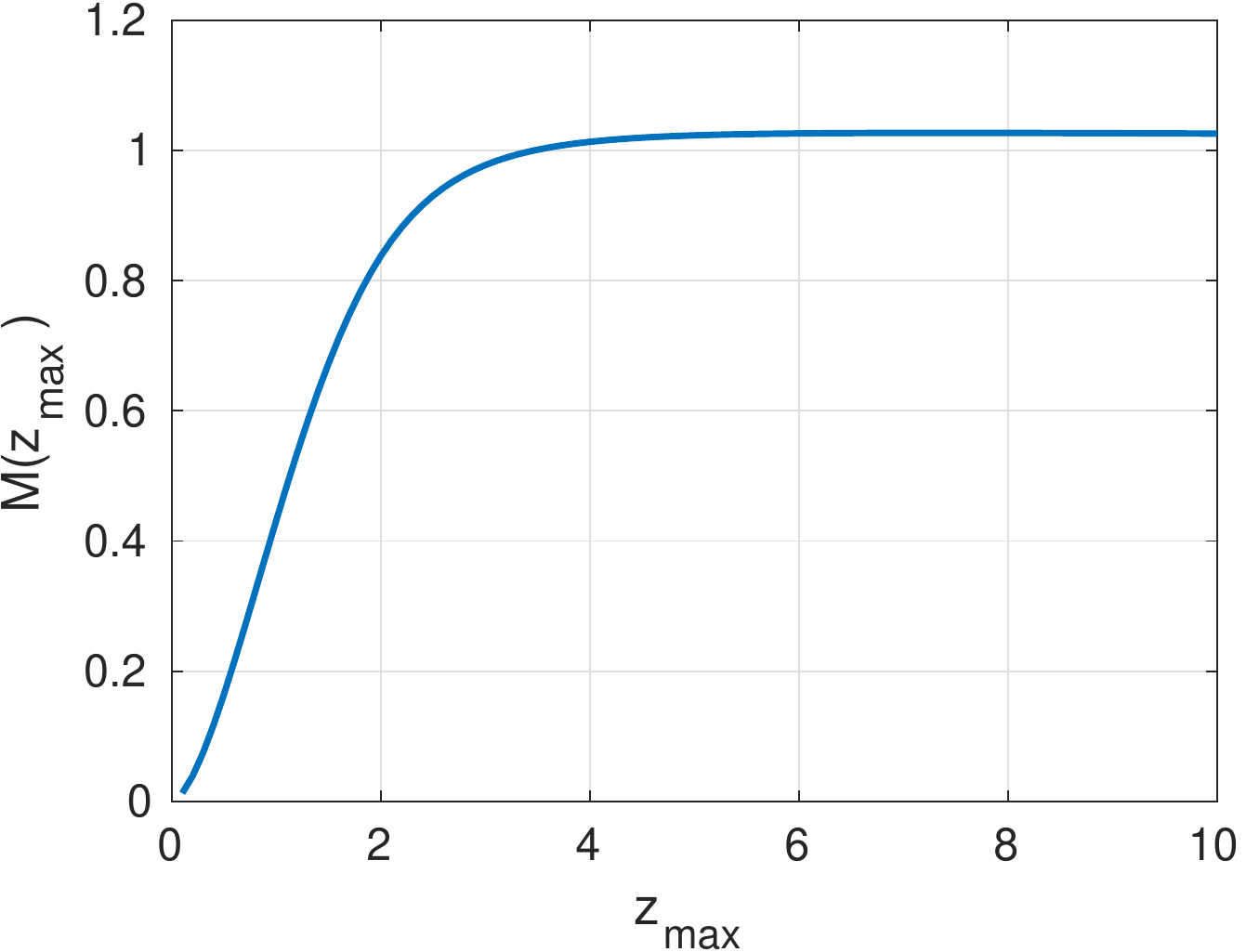}
\caption{The dependence of the integral in Eq.~\ref{eq:frbr} on the redshift.\label{fig:frbrate}}
\end{figure}

After simplifications  and some algebra
we obtain daily magnetar's flare rate of
\begin{equation}
    N_{\rm mag} \approx 10^9  M(z_{\rm max})\, {\rm days}^{-1}\,,
        \label{eq:frbr0}
\end{equation}
where
\begin{equation}
\label{eq:frbr}
\begin{split}
 M(z_{\rm max})&\equiv   \int_0^{z_{\rm max}} \frac{(1+z)^{1/3}}{1+[(1+z)/2.9]^{5.6}}\times \\
 &\frac{1}{\sqrt{(1+z)^3\Omega_{\rm m}+\Omega_{\Lambda}}} I_{\rm D}^{2/3} dz,
 \end{split}
\end{equation}
and
\be
I_{\rm D}\equiv\int_0^z \frac{dz'}{\sqrt{(1+z')^3\Omega_{\rm m}+\Omega_{\Lambda}}}\,.
\ee
For the given choice of parameters the integral saturates at $\sim1$ for $z_{\rm max}>3$ (see in Fig.~\ref{fig:frbrate}). 

Thus, if every magnetar flare produces a radio burst, and the energy of the burst equals  1\% of the total flare energy, then we
expect to see about half-billion events per day above 0.1 Jy. The observed rate is $\lesssim 10^4$ FRBs per day. So,
roughly only one in $\gtrsim 10^5$ magnetar bursts produces a visible FRB. If we account that just about 10\% of magnetars
should have proper conditions in the surrounding medium to generate a flare in $\sim$~GHz range (see in Sec.~\ref{sec:application}), then  FRB generation by magnetar flares should  have
a successive rate of $\lesssim 10^{-4}$.

\section{Discussion and conclusions}

In the estimates above, we have ignored the requirement of small magnetization of flares. Indeed, it was shown that high
frequency maser emission from the RS is possible only if the flare is weakly magnetized.  Typically it is postulated
that magnetars flares are strongly magnetized. However, flares with low magnetization can be
formed along open magnetic field lines at the magnetic pole  in the magnetar magnetosphere. The ratio of the polar caps surface to the magnetar surface
can be estimated as $\eta\sim (R_{\rm M}/R_{\rm lc})/2 \sim 10^{-4} P_{\rm 0}^{-1}$, here $P$ is magnetar spin period;
$R_{\rm lc}$ and \(R_{\rm M}\) are the light cylinder and magnetar radius, respectively.  This simple estimate gives a
result surprisingly close to the required rate of \( \sim 10^{-4}\). This suggests that weakly magnetized flares could
be a very prominent source for production of FRBs through the high frequency maser emission.  This is in correspondence
with the fact that despite many X/$\gamma$-ray flares were registered from SGR 1935+2154 and the source was actively
monitored with radio telescope during its period of activity, just one event was detected simultaneously in radio and in
high energy band (see e.g. \citealt{2021NatAs...5..414K} and references therein).  This burst was much harder than
others in X/$\gamma$-rays \citep{2021NatAs...5..372R}.  {Of course, estimates of the rate made above contains many
  simplifications, so more detailed population synthesis calculations are welcomed.}

\begin{acknowledgments}
The authors appreciate the useful discussions with Sergey Koryagin and Maxim Efremov.
DK acknowledges support by the Russian Science Foundation grant No. 21-12-00416 and by JSPS KAKENHI Grant Numbers 18H03722, 18H05463, and 20H00153.
SP was supported by the Ministry of science and higher education of Russian Federation under the contract 075-15-2020-778 in the framework of the Large scientific projects program within the national project ``Science''.  Study of the conditions required for production of the maser emission at relativsitic shocks was supported by RSF grant No. 21-12-00416. Interpretation of the FRBs in the frameworks of the developed model was supported by the project ``Science'' (contract 075-15-2020-778).
\end{acknowledgments}

%

\vspace{5mm}





\appendix
\section{Synchrotron maser emission}
\label{sec:maser}

Maser (Mircowave Amplification by Stimulated Emission of Radiation) emission allows generating coherent radio flashes
that carry away a per-cent fraction of energy stored in high-energy electrons. For its realization it requires a system
that has inverse population of energy levels with a suitable energy gap. If the underlying emission process is synchrotron
radiation then the maser emission can be generated if the synchrotron self-absorption coefficient,
\be
\label{eq:absorption}
\alpha_\nu = \frac{c^2}{8\pi \nu^2} \int\limits_0^\infty \left[\frac{N(E)}{E^2}\right] \frac{d}{dE}\left[E^2 P_\nu(E)\right] dE\,,
\ee
is negative \citep{1958AuJPh..11..564T}.  Here \(N(E)\) is energy distribution of non-thermal electrons, \(P_\nu(E)\) is
the synchrotron emissivity at frequency \(\nu\) by an electron with energy \(E\), and \(c\) is light speed in vacuum
\citep[see, e.g.,][for a general discussion of the synchrotron self-absorption]{Rybicki&Lightman}. In the range of
frequencies, where the influence of the background plasma is negligible, the synchrotron self-absorption coefficient is
strictly positive, making impossible net amplification of the emission \citep[see, e.g.,][]{1963ARA&A...1..291W}.

The influence of the background plasma appears in the range of electron energies and frequencies where the condition
\(\sqrt{1-\varepsilon(\omega)}\gamma\ll1\) fails. Here \(\varepsilon\) is the dielectric permittivity of the plasma.
For non-relativistic plasma in magnetic field \(\bm B = \bm b B\), the dielectric permittivity is a tensor which has the
following components \citep[see, e.g.,][]{Landau8}
\be
\varepsilon_{\alpha\beta} = \varepsilon_\perp\delta_{\alpha\beta} + (\varepsilon_{||}-\varepsilon_\perp)b_\alpha b_\beta + \imath g e_{\alpha\beta\gamma}b_\gamma\,,
\ee
where \(\delta_{\alpha\beta}\) and \(e_{\alpha\beta\gamma}\) are Kronecker and Levi-Civita tensors, respectively. The
components of dielectric permittivity are determined by the following functions
\be
\varepsilon_\perp=1-\frac{\omega_{p,e}^2}{\omega^2-\omega_{L,e}^2}-\frac{\omega_{p,i}^2}{\omega^2-\omega_{L,i}^2}\,,
\ee
\be
\varepsilon_{||}= 1 - \frac{\omega_{p,e}^2+\omega_{p,i}^2}{\omega^2}\,,
\ee
\be
g=\frac{\omega_{L,e}\omega_{p,e}^2}{\omega(\omega^2-\omega_{L,e}^2)}-\frac{\omega_{L,i}\omega_{p,i}^2}{\omega(\omega^2-\omega_{L,i}^2)}\,.
\ee
The parameters here are plasma frequency
\be
\omega_{p,e}^2=\frac{4\pi n_e e^2}{m_e},\quad\quad \omega_{p,i}^2=\frac{4\pi n_i Z_i^2e^2}{m_i}\,,
\ee
and cyclotron frequency
\be
\omega_{L,e}=\frac{eB}{m_e c},\quad\quad \omega_{L,i}=\frac{eZ_iB}{m_i c}\,,
\ee
where \(m_e\) (\(e\)) and \(m_i\) (\(Z_ie\)) are mass (charge) of electron and ion (or positrons for electron-positron
plasma), respectively. If particles in plasma have relativistic energies, then the dielectric permittivity tensor
depends on the energy and angular distribution of particles \citep[see, e.g.,][]{1984oep.....9.2444A}, which makes challenging  obtaining general analytic
results.  If particles have a narrow energy distribution with energy \(E\), then one can replace in the
  expression for the dielectric permittivity the frequencies with their relativistic counterparts, i.e.
  \(\omega_{L, \cdot }\rightarrow \Omega_{L, \cdot}=(m_ec^2/E)\omega_{L, \cdot} \) and
  \(\omega_{p, \cdot }\rightarrow \Omega_{p, \cdot}=\sqrt{ m_ec^2/E}\,\omega_{p, \cdot}\)
  \citep{1970SvA....13..797S}. \citet{2002ApJ...574..861S} have also shown that this substitution provides an estimate for the permittivity of relativistic plasma
  with accuracy of \(1\%\) if particles have a power-law distribution of index \(2\) above \(E\). We use this approach
  to estimate the frequency at which the emission amplification is possible.

For frequencies significantly exceeding the cyclotron frequency, \(\omega\gg\Omega_{L,e}\), the dielectric permittivity
gets a much simpler form, which is described by a single value
\be
\varepsilon = 1 - \left(\frac{\Omega_{p,e}}{\omega}\right)^2\,,
\ee
where we neglect the contribution from ions (if positrons present, their contribution can be included to the electron one). In this case, the synchrotron emission power is \citep[see, e.g.,][for detail]{1969ARA&A...7..375G}
\be
\begin{split}
P_\nu(E)=& \sqrt{3} \frac{e^2\omega_{\rm L,e}}{ c}\left[1+\left(\frac{\Omega_{\rm p,e}}{2\pi\nu}\right)^2\left(\frac{E}{m_ec^2}\right)^2\right]^{\nicefrac{-1}{2}}\times \\
&\frac{\nu}{\nu_c'}\int\limits_{\nu/\nu_c'}^\infty K_{\nicefrac{5}{3}}(\eta)d\eta\,.
\end{split}
\ee
Here \(K_{\nicefrac{5}{3}}\) is modified Bessel function, and
\be
\nu_c'=\frac{3\omega_{\rm L,e}}{4\pi} \left(\frac{E}{m_ec^2}\right)^2\left[1+\left(\frac{\Omega_{\rm p,e}}{2\pi\nu}\right)^2\left(\frac{E}{m_ec^2}\right)^2\right]^{\nicefrac{-3}{2}}\,.
\ee
\citet{1967JETP...24..381Z} obtained a simple criterion for maser emission \citep[see also in][]{Koryagin2000,Koryagin2006,2017ApJ...842...34W,2018ApJ...864L..12L}: it occurs when the main contribution to the absorption is provided by particles with sufficiently large energy,
\be\label{eq:maser_energy_a}
E >E_{\rm min}= m_ec^2  \frac{2\omega_{L,e} \omega^2}{\Omega_{p,e}^3}\,. 
\ee

\section{Dynamics of the shock formed by intense flare}
\label{sec:MHD}
We consider the following scenario: a powerful flare hits the standing (in the lab frame) shock. In the lab frame, the
flare moves with bulk Lorentz factor \(\Gamma_{\rm fl}\gg1\) and has energy flux of \(F_{\rm fl}\), carried in the form
of bulk motion of cold ejecta and electromagnetic field. This setup is consistent with one of the cases considered by \citet{1976PhFl...19.1130B}, so further details can be found in that paper. 

The ratio of the Poynting flux in the flare to its kinetic
energy flux is \(\sigma_{\rm fl}\). The obstacle against which the flare collides is consists of relativistic hot gas,
with internal energy \(\ve_{\rm pwn}\), magnetic field \(B_{\rm pwn}\). The ratio of energy density of magnetic field to
plasma density is \mbox{\(\sigma_{\rm pwn}\ll1\)} close to the termination shock, \(R\ll10R_{\rm s}\) \citep{1984ApJ...283..694K}. At distances
significantly exceeding the radius of the termination shock, the gas and magnetic pressures are expected to be in
equilibrium.

In the lab reference frame the medium in the PWN moves with a bulk speed of \(c/3\) close to the
termination shock and slows down at larger distances. In what follows we neglect this motion given the large uncertainties of the
flare parameters.

The bulk Lorentz factor of the FS determines the pressure jump:
\be
\ve_{\rm FS} = \frac{8}{3} \ve_{\rm pwn}\Gamma_{\rm FS}^2\frac{v_{\rm FS}}{c}\,,
\ee
where we accounted for the weak magnetization and relativistic equation of state. Typical energy of particles at the FS
down stream is \(m_ec^2\Gamma_{\rm FS}\Gamma_{\rm wind}\), where \(\Gamma_{\rm wind}\) is the bulk Lorentz factor of the
pulsar wind that has blown the PWN.

If the RS moves in the lab frame with a bulk Lorentz factor \(\Gamma_{\rm RS}\), then it is convenient to consider the processes at the RS in its reference frame. The quantities in the RS reference frame we mark with primes. Using the Lorentz transformation we obtain
\be
\Gamma_{\rm fl}'=\frac{\Gamma_{\rm fl}}{2\Gamma_{\rm RS}}\quad\quad{\rm and}\quad\quad F_{\rm fl}'=\frac{F_{\rm fl}}{4\Gamma_{\rm RS}^2}\,.
\ee
Using the jump conditions at the RS, we obtain the internal energy behind the RS:
\be
\ve_{\rm RS} = \frac{8}{3}\frac{F_{\rm fl}}{4\Gamma_{\rm RS}^2c}\,.
\ee
Equating pressures (or internal energies) behind the shocks, we obtain a relation between the shocks' Lorentz factors:
\be
\Gamma_{\rm FS}\Gamma_{\rm RS} = \sqrt{\frac{F_{\rm fl}}{4\ve_{\rm pwn}c}}\,.
\ee
The fluid compression at relativistic shocks is very strong, so we roughly take \(\Gamma_{\rm FS}\approx\Gamma_{\rm RS}=\Gamma\). We therefore obtain 
\be
\Gamma \approx \sqrt[4]{\frac{F_{\rm fl}}{4\ve_{\rm pwn}c}}\approx\frac12\sqrt[4]{\frac{L_{\rm fl}}{L_{\rm sd}}}\,,
\ee
where \(L_{\rm sd}\) is the luminosity of the wind responsible for the formation of the steady nebula.

To estimate the distance to which the flare penetrates into the PWN it is sufficient to estimate the energy carried by the shocked material
\be
E_{\rm s} \approx \Gamma^2 \ve_{\rm pwn} V\approx 4\pi R_s^2\Delta R \Gamma^2 \ve_{\rm pwn}
\ee
(which is valid for \(\Delta R \ll R_s\))
and compare it to the total flare energy
\be
E_{\rm fl} \approx 4\pi R_s^2 F_{\rm fl}\Delta t_{\rm fl}\,.
\ee
Thus, we obtain
\be
\Delta R \approx \sqrt{\frac{4\Delta t^2 F_{\rm fl} c}{\ve_{\rm pwn}}}\,.
\ee
Thus, we obtain an estimate for the flare penetration distance
\be
\Delta R \approx \Delta t_{\rm fl}c\sqrt{\frac{L_{\rm fl} }{L_{\rm sd}}}\ll R_s\,.
\label{eq:dr}
\ee
%





\end{document}